# Absolute gas refractometer without gas-filling and pumping process benefiting from quasi-synthetic wavelength theory

**Jitao Zhang,* Pei Huang, Yan Li, and Haoyun Wei**

*State key Laboratory of Precision Measurement Technology & Instruments, Department of Precision Instruments and Mechanology, Tsinghua University, Beijing 100084, China*

**Corresponding author:* zjt@mail.tsinghua.edu.cn

We present a method to measure the refractive index of gas at 633 nm absolutely, which does not need filling or pumping gas during the measurement. We develop a quasi-synthetic wavelength (QSW) theory by means of the configuration of two-frequency Jamin interferometry and vacuum tubes with specific lengths. With the aid of the QSW theory, we construct a gas refractometer and demonstrate its performance by the measurement of dry air and nitrogen gas at different pressures ranging from 80 kPa to 100 kPa. The results indicate that the refractometer has an uncertainty of better than $1\times10^{-7}$ and a dynamic range of $3.95\times10^{-4}$.

The refractive index of ambient gas is an important input data for optical interferometry and optical design of imaging system where high accuracy and resolution are needed. Since most of the optical interferometers are performed in atmosphere, the refractive index of air has been measured by many methods and currently can be estimated by empirical equations with an uncertainty of a few parts in $10^8$[1-3]. However, the empirical equation is only valid to standard air, and its accuracy is limited by the variation of the components of practical ambient air. Therefore, the direct measurements are usually preferred in place where high accuracy or *in-situ* monitoring is required. The direct methods for measuring gas refractivity absolutely can be categorized by two general methodologies. The first approach is derived from the incremental interferometry, where the phase change of interference fringes along with gas filling from vacuum to ambient or pumping from ambient to vacuum should be recorded continuously[4,5]. This approach is time-consuming and complex since the gas pressure and temperature fluctuation cannot be avoided in the process of filling or pumping. The second approach is measuring gas refractivity absolutely, where the continuous phase-recording, as well as the gas-filling and -pumping, is not necessary [6,7]. Recently, several groups have done successful work through the second approach by means of Fabry-Perot cavity, optical frequency comb, or trapezoidal cavity[8-10]. However, for methods in [8] and [9], Additional measurement in vacuum environment is essential to deduce the final result. And for method in [10], a specific trapezoidal cavity should be fabricated precisely. In this letter, we develop a gas refractometer which can measure refractivity absolutely and directly based on the quasi-synthetic wavelength (QSW) theory. This refractometer has compact configuration, proper accuracy, fast speed, and flexible measurement range, which can be applied to different gas samples.

The optical setup of the gas refractometer is constructed based on the configuration of the Jamin interferometry, which is shown in Fig. 1.

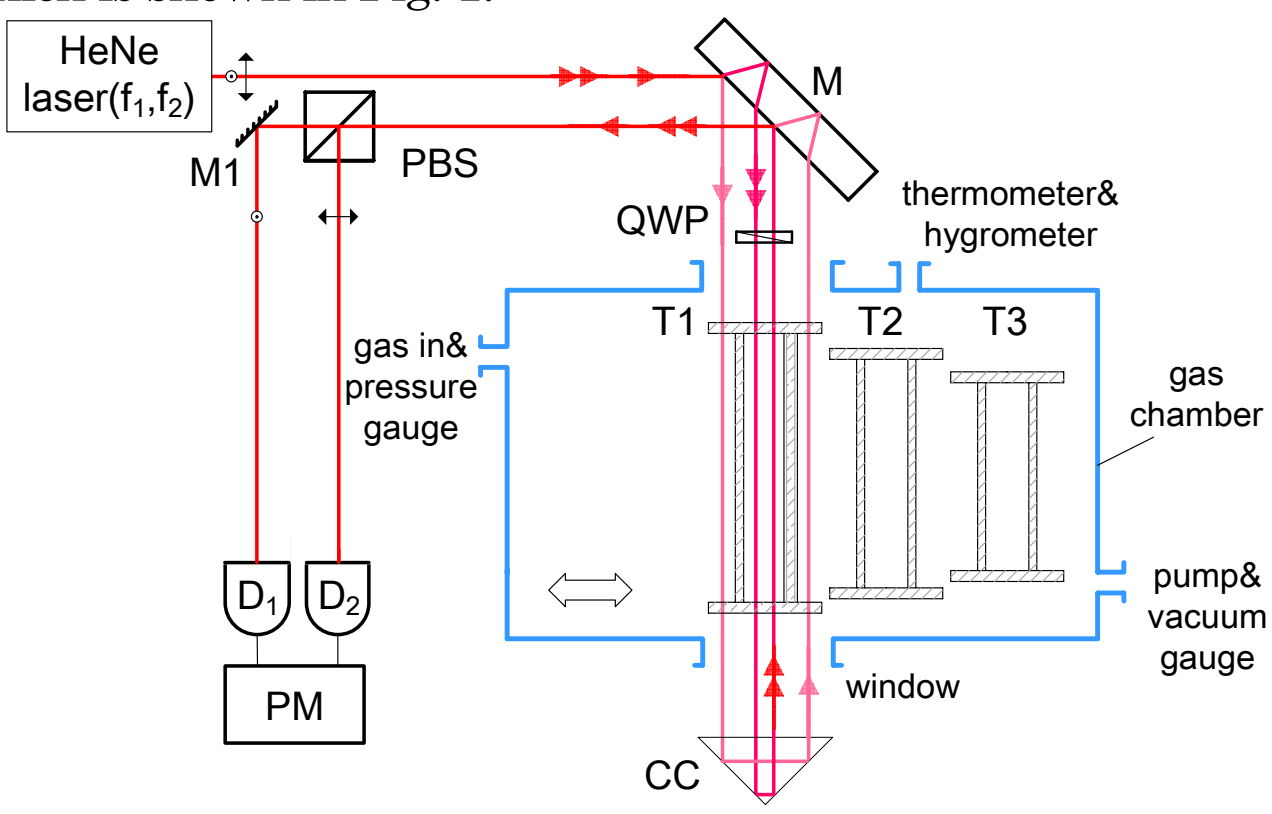

Fig. 1(Color online) Optical setup of the gas refractometer. HeNe laser, laser with two orthogonally linear polarized frequency components $f_1$ and $f_2$; M, plane beam splitter; QWP, quarter waveplate; T1, T2, T3, vacuum tubes; CC, corner cube; PBS, polarized beam splitter; M1, reflector; D1, D2, photodetectors; PM, phase meter.

The polarization direction of two-frequency components $f_1$ and $f_2$ of the HeNe laser is represented by double-headed arrow and concentric circle in Fig. 1, respectively. The M is coated with 50% reflective film on the upper surface and >99% anti-reflection film on the bottom surface. When the incident angle is 45 degree, the HeNe laser can be divided by the M into two parallel beams with similar intensity (shown by double and single arrow in Fig. 1), which then passes through the inner and outer space of the vacuum tube, respectively. After reflected by the CC, these two beam overlay at the upper surface of the M again and interfere. Then, the interfering beam is separated by the PBS into two different polarized components and detected by the $D_1$ and $D_2$. Owing to the QWP, the polarization direction of the inner light is rotated by 90 degree after twice-pass. The end mirrors and the tube of the T1, T2, and T3 are sealed by epoxy resin, and the air of the inner space has been exhausted by a vacuum pump in advance.

The signals $I_1$ and $I_2$ detected by $D_1$ and $D_2$ can be expressed as

$$\left.\begin{aligned} I_1 &= I_{01}\cos\left(2\pi\Delta ft + \Delta\varphi_0 + \Delta\varphi\right) \\ I_2 &= I_{02}\cos\left(2\pi\Delta ft + \Delta\varphi_0 - \Delta\varphi\right) \end{aligned}\right\}, \tag{1}$$

where $I_{01}$ and $I_{02}$ are constant, $\Delta f = f_1 - f_2$ (suppose $f_1 > f_2$ ), $\Delta\varphi_0$ is the difference of initial phases of two frequency components, and $\Delta\varphi$ is the phase shift caused by the refractivity of the gas charged into the chamber. It can be deduced easily from Equ. (1) that the phase difference detected by the PM is $2\Delta\varphi = \frac{1}{2\pi}(N+\varepsilon) = \frac{1}{2\pi}\cdot 4(n-1)L/\lambda$ , which can be written as

$$n - 1 = (N+\varepsilon)\cdot\lambda/4L, \tag{2}$$

where $n$ is the refractive index of charged gas, $N$ and $\varepsilon$ is integral and fractional part of the phase difference $2\Delta\varphi$ normalized by $2\pi$ , $\lambda$ is the wavelength of the HeNe laser, and $L$ is the length of a vacuum tube.

According to the QSW theory, the $\lambda/4L$ in Equ. (2) is denoted by $\lambda_s$ , which is a dimensionless parameter and defined as *quasi-wavelength* (QW) here. Obviously, with different lengths $L$ of vacuum tubes, a series of $\lambda_s$ can be obtained from Equ. (2). The unambiguous measurement range of refractivity in Equ. (2) equals to $\lambda_s$ , which is usually much smaller than routine gas refractivity. Fortunately, thank for the inspiration from the synthetic wavelength interferometry[11,12], this unambiguous range can be expanded by considering measurements with different QWs simultaneously. For instance, with two tubes $L_1$ and $L_2$ , we can deduce two formulas with QWs $\lambda_{s1}$ and $\lambda_{s2}$ according to Equ. (2), and the combination of these formulas can be written as

$$n - 1 = (\Delta N + \Delta\varepsilon)\cdot\Lambda, \tag{3}$$

where $(\Delta N + \Delta\varepsilon)$ is the phase difference between measurements with two QWs, and $\Lambda = 1/\left|\lambda_{s1}^{-1} - \lambda_{s2}^{-1}\right| = \lambda/4\cdot\left|L_1 - L_2\right|$ is defined as QSW here. It is obvious that the range of unambiguity in Equ. (3) is much larger than that in Equ. (2) when the two tubes have close lengths. Similarly, a QSW chain can be organized by choosing several vacuum tubes with proper lengths. For consistency, we call the QWs as zero-order QSW, and higher order QSWs can be generated according to Equ. (3). A case of QSW chain organized by three tubes (with lengths of $L_1$ , $L_2$ and $L_3$ , and suppose $L_1 > L_2 > L_3$ ) has been shown in Fig. 2.

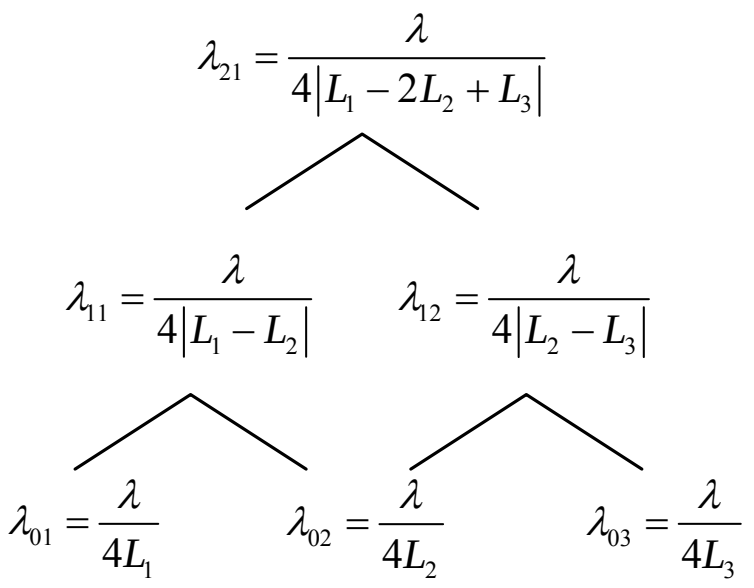

Fig. 2(Color online) A case of QSW chain organized by three vacuum tubes. The symbol with subscript represents QSW, and the first number in the subscript of QSW indicates its order.

If the highest order QSW is larger than the gas refractivity (usually it does), the refractivity deduced by the measured fractional part of phase difference in this order can be served as the initial value to estimate the integral part $N$ in the near lower order QSW. The formula for estimating $N$ is written as

$$N_{j-1} = \mathrm{INT}\left(\frac{(n-1)_j}{\lambda_{j-1}} - \varepsilon_{j-1} + \frac{1}{2}\right), \ j = 1, 2, \ldots, \tag{4}$$

where INT means rounding the element in the bracket to the nearest integer that no more than the element, and the subscript of the parameter indicates the QSW order. Following the same manner above, the gas refractivity can be deduced by the zero-order QSW absolutely in the end. In addition, an essential condition should be always satisfied to make sure the estimated value of $N$ is unambiguous, which is given by

$$\delta(n-1)_j < \frac{1}{2}\lambda_j - \delta(n-1)_{j-1}, \tag{5}$$

where $\delta$ represents the measurement uncertainty of the refractivity in the $j^{\text{th}}$-order QSW.

For experiment, a laser head (Agilent 5517B) is used as the light source, whose vacuum wavelength is 632.991372 nm with an accuracy better than $\pm 1\times 10^{-7}$ (3σ) and frequency difference is about 2.2 MHz. The photodetector (Agilent 10780F) has a bandwidth of about 7 MHz, and the PM (Pretios PT-1313F) has an accuracy of 3.6 degree. The vacuum tubes are made of BK7 glass with 18 mm inner diameter and 5 mm wall thickness, and the inner space of which is pumped down to $10^{-2}$ Pa before sealed. The end mirrors have 42 mm diameter and 5 mm thickness, and are coated by an anti-reflection film with a transmittance of better than 99.8%. According to the QSW theory, the nominal lengths of three vacuum tubes are designed as 165 mm, 158 mm, and 151.5 mm, and the measured lengths after optical fabrication are 164.5864 mm, 157.8515 mm, and 151.5166 mm. Therefore, the measurement range of the refractivity at 633 nm is about $3.95\times 10^{-4}$, which can cover the familiar gases, such as air, nitrogen, and argon, and so on. Three tubes are mounted parallel to each other on a motorized displacement stage, which can transport the tubes into the optical path in length sequence. For a complete measurement of the refractivity, the fractional part of the phase difference is recorded by the PM when each tube is on duty. Then, these phase data are used for calculating phase difference in higher order QSW and deducing the gas refractivity finally. In practice, an additional phase data is also recorded when there is no tube in the optical path, which indicates the phase delay caused by other optical elements except the tubes. This phase delay should be subtracted from each tube's initial phase data before further calculation. Gas pressure and temperature in the chamber need to be measured carefully to provide reference value of the refractivity by the empirical equations. Pressure in the chamber is measured with a digital pressure gauge (Setra Model 470), which has been calibrated with uncertainty of 14 Pa among the range from 80 kPa to 100 kPa. A thermistor sensor (Fluke 1523) is placed close to the current vacuum tube that is in the optical path to monitor the temperature, which has a calibrated uncertainty of 0.01℃. In addition, the chamber should be pumped to vacuum to avoid gas contamination before each target gas is charged, and the residual gas is detected by a capacitance diaphragm gauge (Inficon CDG025D) with calibrated relative uncertainty of 2% below 100 Pa.

The refractivity of dry air and nitrogen gas (purity > 99.9999%) are measured by the gas refractometer. Five times repeated measurements are performed at each interest pressure (roughly from 80 kPa to 100 kPa with a step of 5 kPa) of the target gas filled in the vacuum chamber. We then compared the experimental refractivity to the theoretical refractivity deduced by the empirical formula[3,4,13,14]. For dry air, the reference refractivity is calculated by the modified Edlén equation deduced by Birch[3,4]. For nitrogen gas, the equations deduced by Peck[13] and Zhang[14] are consistent with each other at $1.1\times10^{-8}$ level at 633nm, and the latter is adopted for calculating reference value. The experimental results are shown in Fig.3 &4.

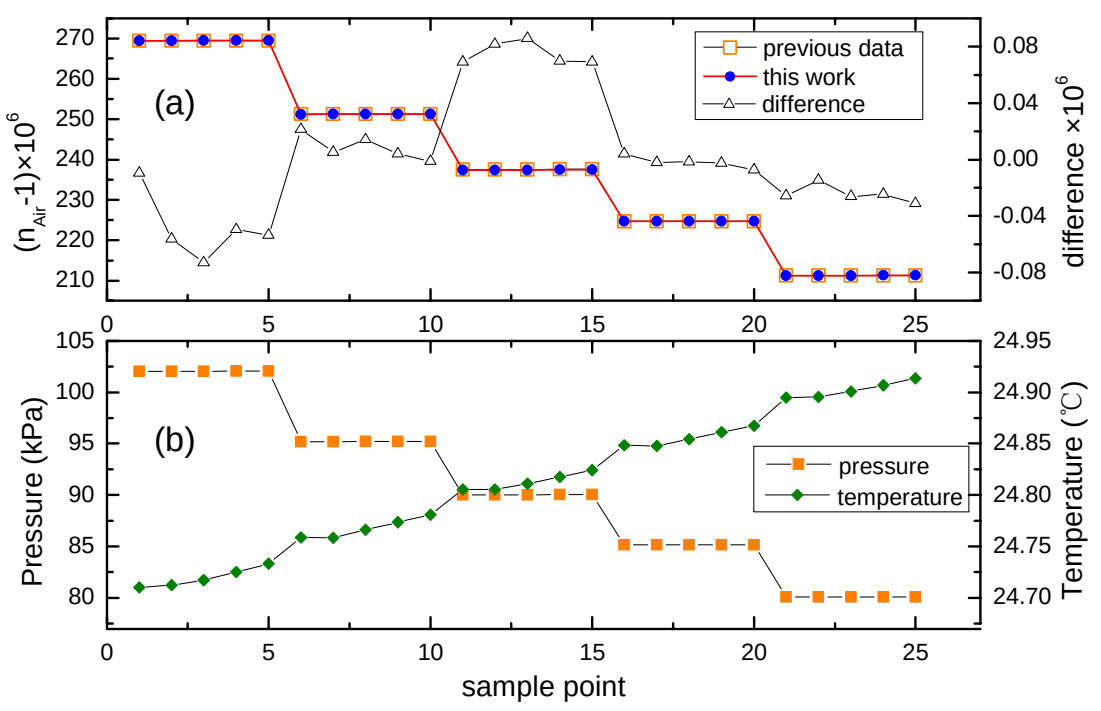

Fig. 3(Color online) Refractivity of dry air at 633 nm. (a) Experimental and theoretical result of dry air. (b) Pressure and temperature data.

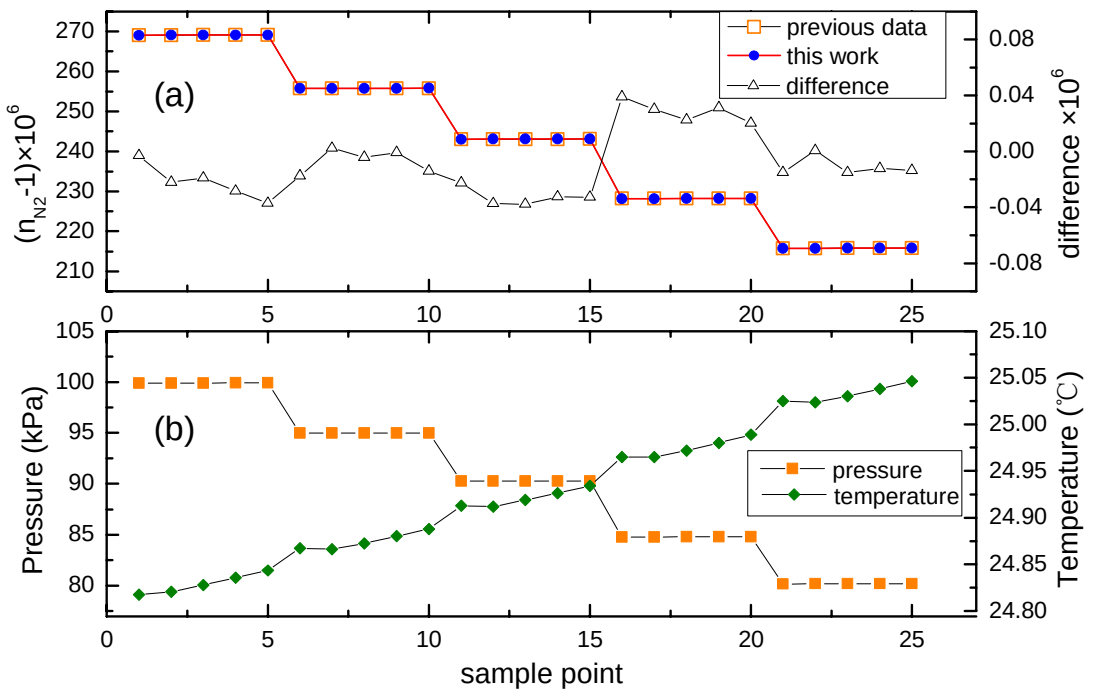

Fig. 4(Color online) Refractivity of nitrogen gas at 633 nm. (a) Experimental result and theoretical result of nitrogen gas. (b) Pressure and temperature data.

It is indicated from Fig. 3&4 that the consistency of the experimental and theoretical data for dry air and nitrogen gas is better than $1\times10^{-7}$ within the routine pressure range. The error sources mainly consist of the lengths of vacuum tubes, phase measurement, and crosstalk of the polarized light. The real length of the tube will vary slightly with the temperature. Since thermal expansion of BK7 glass is $7.1\times10^{-6}$/K, for the longest tube, temperature fluctuation within 10 K will introduce length uncertainty of 11.7 μm. Combining with the length measurement uncertainty of 10 μm, the corresponding uncertainty of the refractive index is $2.8\times10^{-8}$. The measurement uncertainty of the PM is about 3.6°, which is responsible for refractive index uncertainty of $1\times10^{-8}$. In addition, the azimuth error of the QWP's fast axis and the phase delay shifting from exact 90 degree will introduce crosstalk between two orthogonally linear polarized lights, and additional phase will be detected by the PM. The refractive index uncertainty related to the crosstalk is about $1\times10^{-8}$. Overall, the combined uncertainty of the gas refractometer is about $3.1\times10^{-8}$.

In conclusion, we present an absolute method for gas refractivity measurement based on the QSW theory. The principle of the QSW theory is explained in detail, and the performance of this method is demonstrated by the measurement of dry air and nitrogen gas. It takes about 70 s for a complete measurement, and the measurement range can be probably extended by changing the lengths or number of the vacuum tubes. In addition, the uncertainty can be improved by replacing the material of the tubes with ultralow expansion glass and a better PM. Our method can be served for gas refractivity measurement where high accuracy and *in-situ* monitoring are needed.

The authors acknowledge the support of the National Science and Technology Major Project of China.